\documentclass[multphys,vecphys]{svmult}

\usepackage{makeidx}         
\usepackage{graphicx}        
\usepackage{multicol}        
\usepackage[bottom]{footmisc}
\usepackage{url}

\makeindex             

\begin{document}

\title*{Dwarf galaxies in Hickson Compact Groups}
\titlerunning{Dwarf galaxies in HGCs}
\author{Dominik J.\ Bomans\inst{1}\and 
Elvira Krusch\inst{1}\and 
Ralf-J\"urgen Dettmar\inst{1}\and 
Volker M\"uller\inst{2}\and 
Chris Taylor\inst{3}
}
\authorrunning{Bomans et al.}
\institute{
Astronomical Institute of the Ruhr-University Bochum, 
Universit\"atstr. 150, 44780 Bochum, Germany
\texttt{bomans@astro.rub.de, dettmar@astro.rub.de}
\and
Astrophysikalisches Institut Potsdam, An der Sternwarte 16, 
14482 Potsdam, Germany  
\texttt{vmueller@aip.de}
\and 
California State University, Sacramento, 6000 "J" Street
Sacramento, CA 95819-6041, USA 
\texttt{ctaylor@csus.edu}
}
%
%
\maketitle

\begin{abstract}
We observed 5 Hickson Compact Groups with the ESO/MPI 2.2m telescope 
and WFI to investigate the dwarf galaxy content and distribution 
in these galaxy groups.  Our deep imaging and careful selection of the 
candidate galaxies revealed a rich population 
of mainly passively evolving dwarf galaxies, which is spatially much 
more extended than the originally defined Hickson Compact groups. 
The composite luminosity function of the 5 groups shows a bimodal 
structure with a very steep rise in the low luminosity regime. 
The faint end slope is 
close to the predictions of CDM theory for the slope of the Dark Matter 
halo mass function.  
\end{abstract}

\section{Dwarf galaxies in compact groups}
\label{sec:1}
Hickson Compact Groups (HGCs) are tight groups of galaxies selected 
by Hickson on POSS-I plates based galaxy density 
and isolation criteria as described e.g. in \cite{Hickson1997}.
Analysing the environments of HGCs it soon becomes clear, that the cores 
Hickson classified are often not isolated, but embedded into larger 
structures. As an example we found in HGC 16 an spiral galaxy 
with the same redshift, which is located just outside the radius defined 
in Hickson's isolation criterium. 

Searches for dwarf galaxies in HGCs which were conducted up to now 
were either deep, but limited to the area of the compact group itself 
(e.g. \cite{Hunsberger1996}) or wide field, but rather shallow (e.g. 
\cite{Ribeiro1994, Zabludoff2000}.  
We used this as a starting point to investigate the dwarf galaxy content 
of HGCs and to map out the structure of HGCs and their envelopes. 


\section{Observations and sample selection}
\label{sec:2}
We observed nearby HGCs (distance $< 50$ Mpc) using the ESO/MPI 2.2m 
telescope at La Silla with its wide field of view (0.57 
by 0.54 deg), high sensitivity, good spatial sampling (0.24 arcsec per 
pixel), and generally good seeing conditions.  Especially important 
is the good sampling and resolution, which allows us to select 
dwarf galaxy candidates belonging to the HGCs against the background 
of more distant large galaxies.  We observed HGC 16, 19, 30, 31, and 
42 in B and R band, under seeing conditions between 0.8 and 1 arcsec.

\begin{figure}
\centering
\includegraphics[height=7cm]{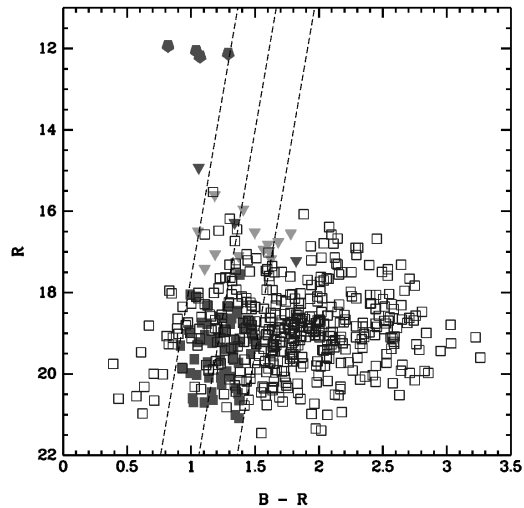}
\caption{Color-magnitude diagram of HGC 16 derived from our photometry. 
Hexagons denote the original HGC 16 members, dark and light triangles 
denote spectroscopic members and non-members from the literature, filled and
open squares denote members and not-members based on our selection criteria.}
\label{fig:1}       
\end{figure}

The data were reduced using IRAF/mscred and some routines developed 
at Bochum.  The final images allowed us to search for galaxies as faint 
as M$_B = -11$ in all observed compact groups. Object detection was done 
using SExtractor, and we set a lower size limit consistent with the 
size of the smalles Local Group dwarf galaxy shifted to the distances 
of the observed HGCs. 
This first selection gave 200-500 dwarf galaxy 
candidates per field. We constructed color-magnitude diagrams (CMDs) for our 
fields and used all available sources for spectroscopic redshifts. An 
example of the resulting CMDs is given in Fig.\ref{fig:1}.  We find very 
few blue dwarf galaxy candidates, while the region of the 
red sequence \cite{Secker1997} is very well populated.  These galaxies are 
therefore candidate dE/dSph galaxies belonging to the HGCs. With the 
good seeing and well sampled images, we can further test this assumption 
similar to \cite{Conselice2002} by classifying all galaxy by eye (select for 
low surface brightness, irregular shape, no spiral arms, and no bulge/disk 
structure). We also analysed the light profiles with surface photometry 
package in IRAF/STSDAS for exponential profiles.

\begin{figure}
\centering
\includegraphics[height=7cm]{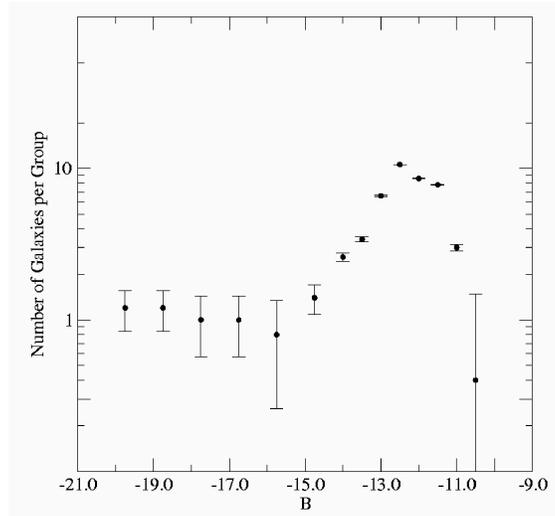}
\caption{Composite luminosity function of all 5 observed HGCs (HGC 16, 19, 
30, 31, and 42).}
\label{fig:2}       
\end{figure}

\section{Dwarf galaxies in compact groups}
\label{sec:3}

The galaxies which passed the selection process described above showed 
a clear concentration towards the HGCs centers, which reassured, that we 
generated samples of high probability dwarf galaxy members of the 
HGCs.  
The spatial extend of these dwarf galaxy population varied from group to 
group, but did not reach the typical background level at a radius of 200 
kpc (limited by the size of our WFI fields) for at least HGC 16, 30, and 42. 

With the assumption of membership for all our final selection dwarf galaxy 
candidates, we generate a composite luminosity function (LF) by converting 
all measurements to a common distance. Given the relatively small differences
in the distance of our target groups this will not induce a large spread 
in observed properties. The resulting LF is plotted in Fig.\ref{fig:2}.

The small number statistics at the bright end precludes an analysis of 
this part and also a formal fit to the whole LF.  What is still obvious 
from the diagram is the steep rise in the dwarf galaxy regime 
(M$_B < -16$ mag) and that the whole LF is not consistent with a single 
Schechter function \cite{Schechter1976}, but implies a bimodal 
LF. The observed decline at luminosities 
fainter than M$_B = -12$ mag is most probably due to the completeness limit 
of our data.  

\begin{figure}
\centering
\includegraphics[height=7cm]{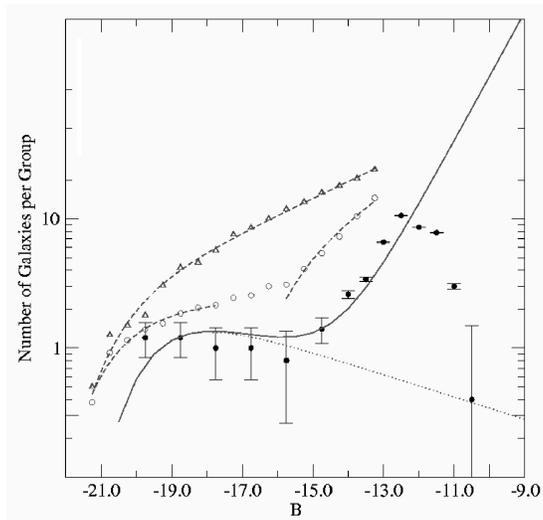}
\caption{Our measured luminosity function (points with error bars) is 
overplotted with the LF for massive (dotted line) and dwarf (solid line) 
passively evolving galaxies from the 2dFRS \cite{Madgwick2002}. Also plotted 
are the measurements for X-ray bright (triangles) and X-ray faint (circles) 
group galaxies from \cite{Miles2004}.}
\label{fig:3}       
\end{figure}

Similar bimodal LFs were recently presented by \cite{Miles2004} for X-ray 
dim groups and are present 
in the LF of passively evolving field galaxies from the 2dFRS 
\cite{Madgwick2002}. Especially the LF derived by \cite{Madgwick2002} 
provides a good agreement for the luminosity of the break in our observed 
LF of HGCs and the extrapolated faint end slope a decent description of our 
data, as shown in Fig.\ref{fig:3}. 

Our data therefore imply a large population of passively evolving dwarf 
galaxies in HGCs.  Such a steep faint end slope dominated by dE/dSph 
galaxies was up to now only found in galaxy clusters, e.g. \cite{Trentham2005}.
In galaxy groups, dE/dSph are generally only found orbiting 
massive galaxies.  Actually, the numbers in the Local Group dwarf galaxies 
seems to be very low compared to the predictions from CDM theory 
(the substructure crisis \cite{Klypin1999, Moore1999}).  In the general 
field, dSph are exceedingly 
rare or even absent. Just recently one object was detected  
which may be the best case yet for a field dSph \cite{Pasquali2005}.  
The steep faint end of the LF we observe in HCGs is hard to understand, 
since the conditions in HGCs are cluster-like only in the very dense cores 
of HGCs, and not in the outer envelopes where we find most of the dwarf 
galaxies.  Here the galaxy densities resemble more those of loose groups.   

\begin{figure}
\centering
\includegraphics[height=7cm]{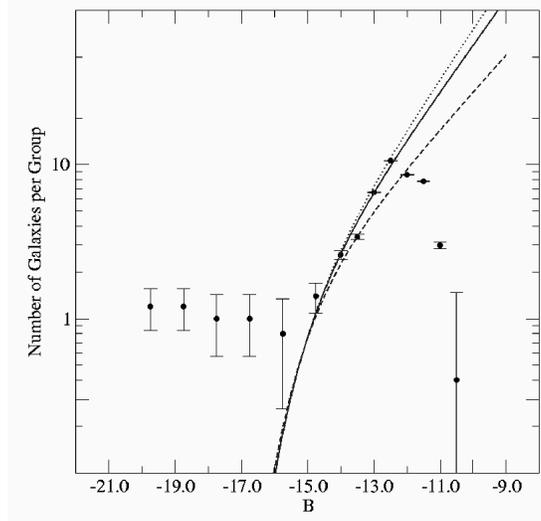}
\caption{Same luminosity function as plotted in Fig.\ref{fig:2} compared with 
faint end slopes of $\alpha = -1.74$ (solid line: best fit to our data), 
$\alpha = -1.6$ and $\alpha = -1.8$ (dashed and dotted lines) overplotted.}
\label{fig:4}       
\end{figure}

The faint end slope of our observed LF is very steep and may add a new 
aspect to the ongoing discussion on the formation, evolution, and survival 
of dwarf galaxies.  The measured LF slope of our HGCs ($\alpha \sim -1.7$) 
is strikingly  similar to the slope of the mass function 
of dark matter halos, see Fig.\ref{fig:4}.  This would imply
that in compact galaxy groups most DM halos would have been populated 
with baryons, in contradiction the results for the Local Group, where 
an overabundance of DM halos without baryons seems to be present.  

We recently got the first VIMOS spectroscopy of the dwarf candidates in two 
of our HCGs, providing redshifts of a large number of the dwarfs. This will 
also allow us to study the dynamics of the dwarf galaxy population and 
the internal properties of the confirmed dwarf members.

%
%
%
%

\printindex
\end{document}